\documentclass[twocolumn,english,amssymb,aps,superscriptaddress,showpacs,amsmath,showkeys,floatfix]{revtex4-2}
\usepackage{amsmath,amssymb,graphicx}
\usepackage{mathptmx}
\usepackage{physics}
\usepackage{color}

\usepackage{geometry}
\usepackage{graphicx}
\usepackage{tikz}

\usepackage[breaklinks]{hyperref}
\usepackage{inputenc}

\begin{document}
	%\title{Dynamics of spatiotemporal solitons from 1D, 2D to 3D: A Diminished Dimension study}
	%\title{Dimensional insights into multidimensional solitons in passive driven cavities with parabolic potentials}
	\title{Multidimensional localized states in externally driven Kerr cavities with a parabolic spatiotemporal potential: a dimensional connection}
	
	\author{Yifan Sun}
	\email{yifan.sun@uniroma1.it}
	\affiliation{DIET, Sapienza University of Rome, Via Eudossiana 18, 00184 Rome, Italy}
	\author{Pedro Parra-Rivas}
	\affiliation{DIET, Sapienza University of Rome, Via Eudossiana 18, 00184 Rome, Italy}
	% \author{Mario Ferraro}
	% \affiliation{DIET, Sapienza University of Rome, Via Eudossiana 18, 00184 Rome, Italy}
	% \author{Wasyhun A. Gemechu}
	% \affiliation{DIET, Sapienza University of Rome, Via Eudossiana 18, 00184 Rome, Italy}
	\author{Fabio Mangini}
	\affiliation{DIET, Sapienza University of Rome, Via Eudossiana 18, 00184 Rome, Italy}
	\author{Stefan Wabnitz}
	\affiliation{DIET, Sapienza University of Rome, Via Eudossiana 18, 00184 Rome, Italy}
	\affiliation{CNR-INO, Istituto Nazionale di Ottica, Via Campi Flegrei 34, 80078 Pozzuoli, Italy}

	\begin{abstract} 
		
In this work, we study the bifurcation structures and the stability of multidimensional localized states within coherently driven Kerr optical cavities with parabolic potentials in 1D, 2D, and 3D systems. Based on symmetric considerations, we transform higher-dimensional models into a single 1D model with a dimension parameter. 
  This transformation not only yields a substantial reduction in computational complexity, but also enables an efficient examination of how dimensionality impacts the system dynamics. In the absence of nonlinearity, we analyze the eigenstates of the linear systems. This allows us to uncover a heightened concentration of the eigenmodes at the center of the potential well, while witnessing a consistent equal spacing among their eigenvalues, as the dimension parameter increases. 
 In the presence of nonlinearity, our findings distinctly reveal that the stability of the localized states diminishes with increasing dimensionality. 
This study offers an approach to tackling high-dimensional problems, shedding light on the fundamental dimensional connections among radially symmetric states across different dimensions, and providing valuable tools for analysis.	
    % Our study provides a valuable tool and enables a deeper physical insight into the mechanisms for the generation of complex high-dimensional dissipative structures.   

% enables us to unveil the fundamental dimensional connection between radially symmetric states with different dimensions.
	\end{abstract}

	\maketitle

	\section{introduction}
 In conservative physical systems, solitons or solitary waves are self-sustained localized wave packets, that maintain their shape unchanged while propagating through nonlinear dispersive or diffractive media. These nonlinear waves have been explored in a plethora of different physical contexts, such as Bose-Einstein condensates, plasmas, hadron matter, gravitation, and optics. \cite{dauxois_physics_2006,malomed_spatiotemporal_2005,kartashov_frontiers_2019,malomed_multidimensional_nodate}. 
 %These states form due to balance between two antagonist effects: nonlinearity and spatial coupling (e.g., dispersion, diffraction, diffusion). 
 A completely different class of localized waves appears in open or dissipative systems, 
 %which are far from thermodynamic equilibrium where the spatial coupling and nonlinear equilibrium are complemented
 which experience a continuous energy exchange with the surrounding media. These localized dissipative states, hereafter LS, are also known as dissipative solitons \cite{akhmediev2008dissipative}.
In dissipative optical systems, LS are under the balance between gain and loss, and
manifest in systems of different dimensionality: e.g., 1D temporal LS involve a balance between nonlinearity and dispersion, while 2D LS form in the presence of nonlinearity and diffraction.
The theoretical prediction of the emergence of 2D spatial LS \cite{Lugiato1987} was experimentally confirmed using a large-area vertical cavity surface emitting laser with a coherent holding field \cite{Barland2002}. 
In parallel, the emergence of temporal cavity solitons, the counterparts of spatial cavity solitons, was predicted by substituting diffraction with dispersion \cite{haelterman_dissipative_1992,Wabnitz1993}. Subsequent experimental confirmations of temporal dissipative Kerr solitons in fiber optics \cite{Leo2010} and microresonators \cite{Herr2014} highlighted their connection with the generation of coherent optical frequency combs. These combs unlock a new era of technological possibilities, including optical buffering, optical clocks, dual-comb spectroscopy, and frequency synthesizers.\cite{Leo2010,Kippenberg2018}. 
	
Diffraction and dispersive effects may occur simultaneously, leading to the localization of light in space and time, and consequently giving rise to the formation of spatiotemporal 3D solitons
\cite{kivshar_optical_2003,malomed_multidimensional_nodate,kartashov_frontiers_2019}. These spatiotemporal localized waves, also called light bullets \cite{silberberg_collapse_1990, javaloyes2016cavity}, have been experimentally observed in both conservative \cite{minardi_three-dimensional_2010,Renninger2013,panagiotopoulos_super_2015}, and dissipative systems, such as spatiotemporal mode-locked lasers \cite{Wright2017} and multimode Fabry-Perot microresonators \cite{Nie2022}. 
% Furthermore, experimental evidence supports their generation in multimode Fabry-Perot microresonators \cite{Nie2022}. 

\begin{figure}[t]
\centering
\includegraphics[width=1\columnwidth]{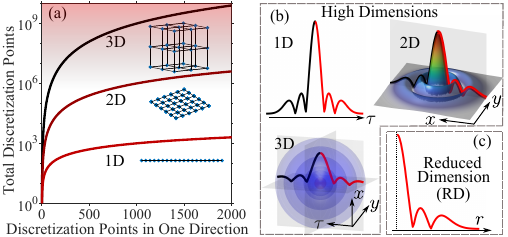}
\caption{
    (a) Total number of sampling points required for physical models in 1D, 2D, and 3D models depend on the sampling points in a single direction. 
    The shallow red region represents computational loads that are extremely heavy for a typical personal computer.
    Panel (b) shows radial symmetric multidimensional LS in 1D, 2D, and 3D models, featuring highlighted radial profiles represented by red curves. In (c), multidimensional LS in (b) can be analyzed in terms of their radial profile, by leveraging on their symmetry.
}
\label{fig1}
\end{figure}

 Nevertheless, theoretical studies of 3D dissipative solitons \cite{Rivas2023OC,Raghavan2000Agrawal,Wright2020,Mayteevarunyoo2019,Javaloyes2016a,Gurevich2017,Gopalakrishnan2021,Kalashnikov_2022,Sun2023bullets} still face many fundamental and practical challenges. From a fundamental perspective, an important problem is the fragility of multidimensional states against any perturbation \cite{kartashov_frontiers_2019}. A typical example of such fragility is the spatial or spatiotemporal wave collapse, that is commonly experienced by multidimensional solitons in optical materials with Kerr nonlinearity \cite{kartashov_frontiers_2019,berge_wave_1998-1,kuznetsov_bifurcations_2011,Rivas2023OC}.
	To arrest, at least partially, these catastrophic instabilities, various strategies have been proposed, including dynamical regularization of collapse \cite{bree_regularizing_2017}, the use of saturable, nonlocal, and competing nonlinearities \cite{Akhmediev1993, edmundson_robust_1992, bang_collapse_2002, mihalache_three-dimensional_2006,desyatnikov_three-dimensional_2000,mihalache_stable_2002}, rapid longitudinal variations of the material parameters \cite{Matuszewski2004}, the use of both static and twisted optical lattices \cite{Aceves1994,aceves1995energy, morsch_dynamics_2006, christodoulides_discretizing_2003, kartashov_soliton_nodate} and other wave confinement methods \cite{milian_robust_2019,raghavan_spatiotemporal_2000,shtyrina_coexistence_2018}. 
	From the modeling perspective, a key challenge arises when considering the high-demanding computational load for the study of multidimensional LS.
 % , specifically when solving initial value problems.
	When the dimensionality increases, the computational complexity grows significantly larger, posing a bottleneck for numerical investigations of high-dimensional solitons. 
    Typically, the workload grows quadratically for 2D problems and cubically for 3D problems, yielding an insufficient number of points in the discrete grids. For instance, when employing a thousand discretization points in 1D corresponds to a million points in 2D, and a billion samples in 3D [see Fig.~\ref{fig1}(a)]. 
	Thereby, in high-dimensional problems, the demand for an excessively large number of grid points poses a key obstacle, since for practical reasons one is forced to use very low-resolution meshes, which may result in significant numerical errors.
	Moreover, the essential question that we try to address here is: how and why dimensionality affects the stability of LS?

	In this work, we answer this question by performing an extensive bifurcation analysis of radially symmetric LS, in the context of externally driven Kerr cavities with a spatiotemporal parabolic confining potential \cite{Sun2023bullets}. When considering this symmetry, one may reduce the description of multidimensional LS [see Fig.\ref{fig1}(b)] to its 1D radial profile [see Fig.\ref{fig1}(c)], which greatly facilitates the analysis across an arbitrary number of dimensions. 
	By doing so, we solve the issue of the heavy numerical computation load for simulating the dynamics of highly dimensional states. Moreover, we are able to unveil why the latter are much more fragile than their 1D counterparts. Clearly, the dimensional reduction based on the radial symmetry inherently removes the possibility of studying other types of more complex states, which may emerge from the radial ones \cite{lloyd_localized_2008,avitabile_snake_2010,gopalakrishnan_dissipative_2021}. On the one hand, this has the benefit of considerably facilitating the analysis; however, it leaves out the possible emergence of radially asymmetric states in the original systems. In any case, our procedure has the merit that it enables us to unveil the fundamental dimensional connection between radially symmetric states with different dimensions.
	
	This article is organized as follows. In Section~\ref{sec:2} we introduce the mathematical nonlinear model describing the evolution of the optical field in the cavity. Section~\ref{sec:3} explores the properties of the eigenmodes, which are associated with the linear problem. Later, in Section~\ref{sec:4} we study how, in the nonlinear regime, such modes lead to nonlinear LS. The bifurcation and stability analysis of these states is presented in Section~\ref{sec:5}, where the dimensionality implications are carefully addressed. Finally, in Section~\ref{sec:6} we present a short discussion and draw our main conclusions. 
	
	% \section{The radial driven dissipative Gross-Pitaevskii model}\label{sec:2}
	\section{The radial model}\label{sec:2}
	In the mean-field approximation, the spatiotemporal field dynamics in externally driven Kerr cavities, in the presence of diffraction and/or anomalous dispersion, can be described in terms of the driven dissipative Gross-Pitaevskii equation 
	\begin{equation} 
		\frac{\partial A}{\partial t} =\mathrm{i}\left[\nabla^2 - \mathbf{r}^2\right]A 
		+ \mathrm{i}|A|^2A  - (1+\mathrm{i}\delta)A+ P,
		\label{eq1}
	\end{equation}
	where $A(\mathbf{r},t)$ is the electric field envelope, $\mathbf{r}\in {\rm R}^D$ with $D$ being the dimension, $t$ is time, $\delta$ is the cavity phase detuning, $P$ is the intensity of the external driving field or pump, and the losses are normalized to $-1$. 
    % The term $r^2$, with $r\equiv |\mathbf{r}|$, represents the parabolic potential, and $\nabla^2$ is the $D$-dimensional Laplacian. 
    $\nabla^2$ is the $D$-dimensional Laplacian and the term $\mathbf{r}^2$ represents the parabolic potential $\mathbf{r}^2=\tau^2$ in 1D, or $\mathbf{r}^2=x^2+y^2$ in 2D or $\mathbf{r}^2=x^2+y^2+\tau^2$ in 3D. 
    Note that special cases of Eq.~(\ref{eq1}) describe the dynamics of temporal states $A(\mathbf{r},t)=A(\tau,t)$ in dispersive cavities when $D=1$, of spatial structures in diffractive cavities when $D=2$ [i.e., $A(\mathbf{r},t)=A(x,y,t)$], and for $D=3$ of spatiotemporal states $A(\mathbf{r},t)=A(\tau,x,y,t)$, with diffraction and dispersion acting simultaneously.

	By polar or spherical coordinate transformations, radially symmetric solutions $A(\mathbf{r},t)=A(r,t)$ of Eq.~(\ref{eq1}) in the region $(0,R)$ obey
	\begin{equation} 
		\frac{\partial A}{\partial t} =\mathrm{i}\left[\frac{\partial^2 A}{\partial r^2}+\frac{D-1}{r}\frac{\partial A}{\partial r} - r^2\right]A 
		+ \mathrm{i}|A|^2A  - (1+\mathrm{i}\delta)A+ P,
		\label{eq_main_321}
	\end{equation}
	together with the boundary condition $\partial_r A(r)|_{r=0,\,\mathrm{or},\,r=R}=0$.
%	{\color{green}boundary conditions [ADD BOUNARY CONDITIONS] ...}
	Here, the term $r^{-1}(D-1)\partial_r A$, arising from the $D$-dimensional Laplacian $\nabla^2A$, is the pivotal element for distinguishing the dimensionality of the problem. For $D>1$, this term introduces an effective position-dependent field flow: in its presence, the field experiences an elevated flow rate when $r\rightarrow{0}$, which leads to an effective centripetal force, weighted by $D$, compelling the field to concentrate towards the center $r=0$.
	
	Equation~(\ref{eq1}) has been previously used to study 1D temporal LS in dispersive Kerr cavities with external phase modulation \cite{Sun2022OL,Sun2023chaos,Sun2023transition}, and for analyzing the stabilization of high-order spherical (3D) light bullets, when diffraction and dispersion act simultaneously in the presence of a spatiotemporal potential \cite{Sun2023bullets}. Similarly, a diffractive cavity with a spatial parabolic inhomogeneity can be described by Eq.~(\ref{eq_main_321}) when $D=2$. 
	Building on these results, now we consider $D$ as a continuous parameter. This permits us to explore the homotopic connection between radially symmetric LS across different dimensions, as well as the impact of the dimensionality on the stability of LS.

	%Leveraging the inherent symmetry of solutions in these systems, we reduce high-dimensional models to a reduced-dimensional form, enabling the convenient study of dimensionality's impact on solitons through a single dimensional parameter.

	%{\color{red}Model.}
	%We investigate three driven damp nonlinear Schrodinger equation models featuring parabolic potentials operating in different dimensions.
	%By implementing polar and spherical coordinate transformations, we can rewrite the field amplitude $A_1(\tau)$ in 1D, $A_2(x,y)$ in 2D, and $A_3(x,y,\tau)$ in 3D into their radial form $A(r)$, by considering azimuthal and elevation directions as homogeneous [see details in Appendix \ref{appen:model_reduction}]. 
	%This assumption  will be discussed later.
	%Therefore, the 1D model in Ref.~\cite{Sun2023chaos}, the 3D model in Ref.~\cite{Sun2023bullets}, and a similar model in the 2D case, can all be reduced to a unified, normalized equation:
	
% {\color{red}	Physically, the radial localization around $r=0$ is prompted by two main reasons. 
% 	First, the presence of a coherent and spatially uniform pump $P$ effectively suppresses asymmetric field modes. This is because the energy pumping rate for each mode is intimately linked with the transverse integral of the modal profiles \cite{Sun2023chaos}, thereby favoring the dynamics dominated by symmetric solutions.
% 	Second, the presence of a parabolic potential breaks the translational invariance of LS, effectively forcing the field to be distributed around the center of the potential well.
%  }

Physically, the radial localization around $r=0$ is prompted by two main reasons. 
First, the presence of a parabolic potential breaks the translational invariance of LS, effectively forcing the field to be distributed around the center of the potential well.
Second, the presence of a coherent and spatially uniform pump $P$ effectively suppresses asymmetric field modes. This is because the energy pumping rate for each mode is intimately linked with the transverse integral of the modal profiles \cite{Sun2023chaos}, thereby favoring the dynamics dominated by symmetric solutions.

	%%%%%%%%%%%%%%%%%%%
	%Q1$$$$$$$$$$$$$$$$$$
	%{\color{red}in the low pump power regime.} 

	All of the results presented here have been mutually validated through path-continuation (homotopic) techniques by using pde2path \cite{Uecker2014a}, as well as by means of full dimensional direct numerical simulations (DNS). Similarly, the LS stability was evaluated through both spectral analysis and DNS.
	
	\section{Linear eigenmodes}\label{sec:3}
	Building upon the foundation laid by Eq.~(\ref{eq_main_321}), let us start by analyzing the linear eigenvalue problem
	\begin{equation}
		\hat{H}_D \psi_{D,n} = \mathrm{i} \delta_{D,n} \psi_{D,n},   
	\end{equation}
	where $\hat{H}_D$ is the linear operator  
	\begin{equation}
        \hat{H}_D\equiv\mathrm{i}\left(\frac{D-1}{r}\frac{\partial }{\partial r} + \frac{\partial^2}{\partial r^2} - r^2\right),
	\end{equation}
	and $\psi_{D,n}$ and $\delta_{D,n}$ are the linear eigenmodes and eigenvalues, parametrized by the dimension parameter $D$ and the mode number $n$. 
	
\begin{figure}[!t]
		\centering
		\includegraphics[width=\columnwidth]{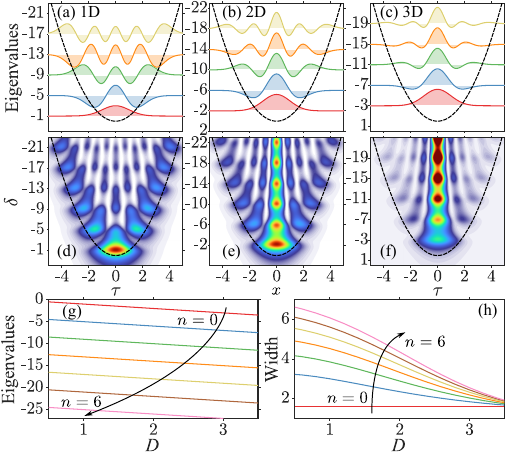}
		\caption{
			Eigenmode profile comparison of (a) 1D HG modes, (b) 2D LG modes, (c) 3D spherical modes. 
			Stable field intensity distribution $|A(\delta,\tau)|^2$ for 1D in (d), $|A(\delta,x,y=0)|^2$ for 2D in (e), $|A(\delta,\tau,x=0,y=0)|^2$ for 3D in (f) vs. $\delta$ in the absence of Kerr effect.
            The colormap illustrates a gradient of intensities, transitioning from white to blue, then green, and reaching red.
			Eigenvalues in (g) and amplitude-weighted mode width in (h) as a function of mode order $n$. 
		}
		\label{fig_eig}
	\end{figure}

	%This operator is crucial for determining eigenmodes $\psi_{D,n}$ and their associated eigenvalues $\delta_{D,n}$ across different dimension $D$. The relationship is expressed as $\hat{H}_D \psi_{D,n} = \mathrm{i} \delta_{D,n} \psi_{D,n}$, with $n$ being the mode order.
	%Note that this analysis is tailored specifically to the radial components of the field, thus limiting its scope to symmetric modes.
	In order to visually represent these radial modes, we mirror their profiles to negative $r$-values, so as to construct full 1D profiles. These eigenmode profiles in Fig.~\ref{fig_eig}(a-c) are vertically shifted according to their respective eigenvalues for the 1D, 2D, and 3D scenarios, respectively. 
	Even though the dimension of a physical system is discrete, we analyze the modification of the eigenvalues $\delta_{D,n}$ by continuously varying the dimension parameter $D$ [see Fig.~\ref{fig_eig}(g)].
	
	As it can be seen, the eigenvalues remain equidistant for different $D$. In other words, increasing the dimension does not affect the spacing between eigenvalues; instead, it uniformly shifts upwards all the eigenvalues in the potential as $D$ increases. 
	The mode separation $g$ depends on the coefficients of diffraction ($C_1$) and dispersion ($C_2$) through the expression $g=2\sqrt{4C_1/C_2}$. Note that, in our normalized Eq.(\ref{eq_main_321}), $C_1=C_2=1$. Thus, the relation governing these eigenvalues reads
	\begin{equation}
		\delta_{D,n} = -[g(n+1/2)+D].
	\end{equation}
	The mode profiles in Fig.~\ref{fig_eig}(a-c) exhibit distinct characteristics: they manifest as symmetric Hermite-Gaussian (HG) modes in 1D, and as symmetric Laguerre-Gaussian (LG) modes in 2D, and 3D spherical modes, respectively. 
	
	As the mode order $n$ increases, the spatial profile of the eigenmodes grows wider within the potential. However, for higher dimensions, the modes become more tightly localized around the center of the potential well. For high-order modes, the highest intensity peaks are located at the potential boundaries in 1D, but they shift toward the center, and the central peak becomes dominant for 2D and 3D.
	This observation can be quantified through the calculation of the amplitude-weighted modal width 
	\begin{equation}
		w(D,n)=\left. 2\int |\psi_{D,n}(r)|r\dd r \middle/ \int |\psi_{D,n}(r)|\dd r \right.,
	\end{equation}
	which is shown in Fig.~\ref{fig_eig}(h). The concentration of the modal profiles towards the center of the potential is a pivotal dimension-related signature.

	The resonant modes for different eigenvalues are verified by numerically solving their associated equations within original number of dimensions (Eq.~(\ref{eq1})), in the absence of Kerr nonlinearity $\mathrm{i}|A|^2A$ [see Figs.~\ref{fig_eig}(d, e, f)].
	This permits to depict the distribution of the corresponding fields as $|A(r)|^2=|A(\tau)|^2$ in 1D, $|A(r)|^2=|A(x, y=0)|^2$ in 2D, and $|A(r)|^2=|A(\tau, x=0, y=0)|^2$ in 3D, for a fixed value of $P=1.3$. 
    These distributions strikingly resemble the eigenmodes of Figs.~\ref{fig_eig}(a, b, c), which confirms their appropriateness in describing the resonant modes of the linear cavity. 
    For other pump values, the fields exhibit analogous distributions as they correspond to solutions in linear scenarios.

	\section{Bifurcation structure of nonlinear localized states and dimensional connection}\label{sec:4}
	
\begin{figure}[t]
	\centering
	\includegraphics[width=\columnwidth]{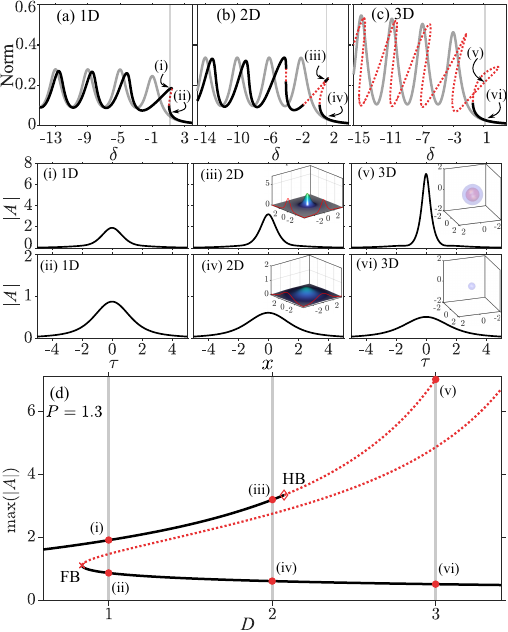}			
	\caption{
		(a-c) Comparison of bifurcation diagrams of the field norm vs. $\delta$ in the absence (gray curve) and in the presence (black solid and red dashed curve) of Kerr nonlinearity for different dimensions: 1D in (a), 2D in (b), and 3D in (c), when $P=1.3$. Vertical lines in (a-c) correspond to $\delta=1.2$. 
		(i-vi) Individual solutions for the field amplitude distribution $|A|$, corresponding to the solution marked (i-vi) in the bifurcation diagrams (a-d).
		(d) Bifurcation diagram of the maximum amplitude vs. the dimension parameter $D$, for $P=1.3$ and $\delta=1.2$.
	}
	\label{fig3_non}
	\end{figure}

	In order to elucidate the influence of Kerr nonlinearity on the system dynamics, let us compare the changes of linear and nonlinear LS when scanning the cavity detuning $\delta$ for a fixed pump amplitude $P=1.3$.
	To do that, we may compare the variation of the norm $N = \sqrt{\int |A(r)|^2 \dd r}/L$ associated with the 1D radial profiles with detuning $\delta$,  where $L=30$ is twice the radial domain size. These results are illustrated in the bifurcation diagrams of Figs.\ref{fig3_non}(a,b,c) for $D=1,2,3$.
%	, which have been computed trough path-continuation techniques \cite{uecker_numerical_2021}.
	
	%----------------------------
	
	In the absence of Kerr nonlinearity [see gray curves in Figs.~\ref{fig3_non}(a,b,c)], all solutions are stable, aligning linear resonance peaks precisely with their eigenvalues [Figs.~\ref{fig_eig}(a,b,c,g) for $D=1,\,2,\,3$]. 
	The introduction of Kerr nonlinearity shifts the resonance peaks in Figs.\ref{fig3_non}(a,b,c) towards positive $\delta$-values, leading to both stable (depicted by black solid curves) and unstable solutions (depicted by red dashed curves).
	This shift is due to the fact that modes resonate with an effective nonlinear cavity detuning $\mathrm{i}(|A|^2-\delta_{D,n})$ [see Eq.~(\ref{eq_main_321})], owing to the intensity-dependent phase shift $\mathrm{i}|A|^2$. 
	In order to maintain the effective detuning nearly constant at each resonance, the linear detuning $\delta$ undergoes a substantial adjustment with increasing values of the intensity $|A|^2$. In order words, higher intensities leads to larger values of $\delta$, leading to the bending of resonance peaks and to the formation of fold bifurcations (FB).

	Despite sharing an identical pump amplitude $P=1.3$, there are noteworthy differences among the scenarios observed for different dimensions. 
	In 1D [see Fig.~\ref{fig3_non}(a)], a single pair of fold bifurcation points emerges at the first resonance peak, and a slight tilting is observed. This leads to the coexistence of two different LS for the same $\delta$-value (i.e., bistability). Two examples of such states are depicted in panels Figs.~\ref{fig3_non}(i,ii) for $\delta=1.2$ [see labels in Fig.~\ref{fig3_non}(a)].
	
	%To visualize 1D field amplitude profiles $|A|$, we have provided 
	%the solutions, labeled (i,ii) in Fig.~\ref{fig3_non}(a) for $\delta=1.2$, are depicted in Fig.\ref{fig3_non}(i,ii). 
	
	In 2D [see Fig.~\ref{fig3_non}(b)], two pairs of FB points appear around the first two resonance peaks. 
	Between these peaks, 2D breathers [see the dashed curve] appear.
%	 (not shown here) {\color{blue}[see Appendix \ref{appen:compare_DNS_conti_eigenvalue} for more details]}.
	Coexisting (i.e., bistable) LS are shown in Figs.~\ref{fig3_non}(iii,iv) for $\delta=1.2$. A reconstruction of the 2D field $A(x,y)$ is obtained from the radial profiles and shown in Figs.~\ref{fig3_non}(iii,iv) inset.
	%are shown in Fig.\ref{fig3_non}(iii,iv).

	Notably, in 3D all solutions become unstable, except those on the lower branch of the first resonance peak; moreover, much stronger resonance tilts are observed in this case.
	Solution (v) [see Fig.~\ref{fig3_non}(v)] becomes unstable, while solution (vi) [see Fig.~\ref{fig3_non}(vi)] remains stable for $\delta=1.2$, as shown in Figs.~\ref{fig3_non}(c,v,vi). In this case, we also display the 3D reconstruction of the field $A(x,y,\tau)$ (see insets in Figs.~\ref{fig3_non}(v,vi)). 
	To validate the results shown in Figs.~\ref{fig3_non}(i-vi), including their linear stability, we have performed DNS in the original 1D, 2D and 3D models.
	%, and later project the radial profiles of 2D (iii,vi) and 3D (v,vi) onto their respective full 2D and 3D coordinates, constructing $A(x,y)$ and $A(x,y,\tau)$. These full formats are depicted in corresponding subpanels in Fig.\ref{fig3_non}(iii-vi). We confirm the solutions their full dimensions and solution stability in Fig.~\ref{fig3_non}(a,b,c) by running DNS.

	Building upon this insight, we explored the modifications of the bifurcation structures across a continuous variation of the dimensionality of the system, by harnessing the dimension parameter $D$ as a continuous variable,
 % \cite{mccalla_snaking_2010}, 
    even though the dimension parameter is inherently a discrete quantity. 
	To examine the validity of this approach, we chose a pair of LS solutions within the bistable region, for fixed values of $P=1.3$ and $\delta=1.2$ [see the gray vertical lines in Figs.~\ref{fig3_non}(a-c)], and performed their path-continuation in $D$. The outcome of these computations is shown in 
	Fig.~\ref{fig3_non}(d), where we plot the modification of the maximal norm of the radial profile $|A(r)|$ as a function of $D$. 
	In this diagram, the solutions (i-vi) at $D=1,2,3$ correspond exactly to the solutions (i-vi) indicated with vertical gray lines in Figs.~\ref{fig3_non}(a-c) and Figs.~\ref{fig3_non}(i-vi), respectively. 
	In other words, one may effectively investigate the evolution of multidimensional solitons across their dimensionality, by simply exploiting a single dimension parameter $D$.
	
	The spectral (i.e., linear) stability analysis carried out for this diagram, limited to the presence of radial perturbations, indicates that 
	the LS states (ii,iv,vi) [see bottom branch in Fig.\ref{fig3_non}(d)] are stable. In contrast, in the upper branch, only the 1D and 2D LS are stable. After crossing $D=2$, a Hopf bifurcation (HB) emerges, which destabilizes this branch.
	%point, indicating higher dimensions bring more instability.
	It is worth noting that despite obtaining solutions (i, iii, v) under identical parameters $\delta$ and $P$, their field amplitude peaks exhibit a substantial increase with dimensionality. 
	In Fig.\ref{fig3_non}(d), we depict the maximum amplitude as $\mathrm{max}(|A|)$, rather than intensity $\mathrm{max}(|A|^2)$ for clarity: the resulting intensity difference is notably pronounced. 
	This suggests the presence of substantial intensity-dependent phase modification (i.e., owing to nonlinearity) in the higher-dimensional systems.
	Consequently, one can anticipate that in these systems reducing the pump amplitude in these systems may enhance the stability of the LS.

	\section{Bifurcation analysis through dimensions} \label{sec:5}

 \begin{figure*}[t]
		\centering
		\includegraphics[scale=1]{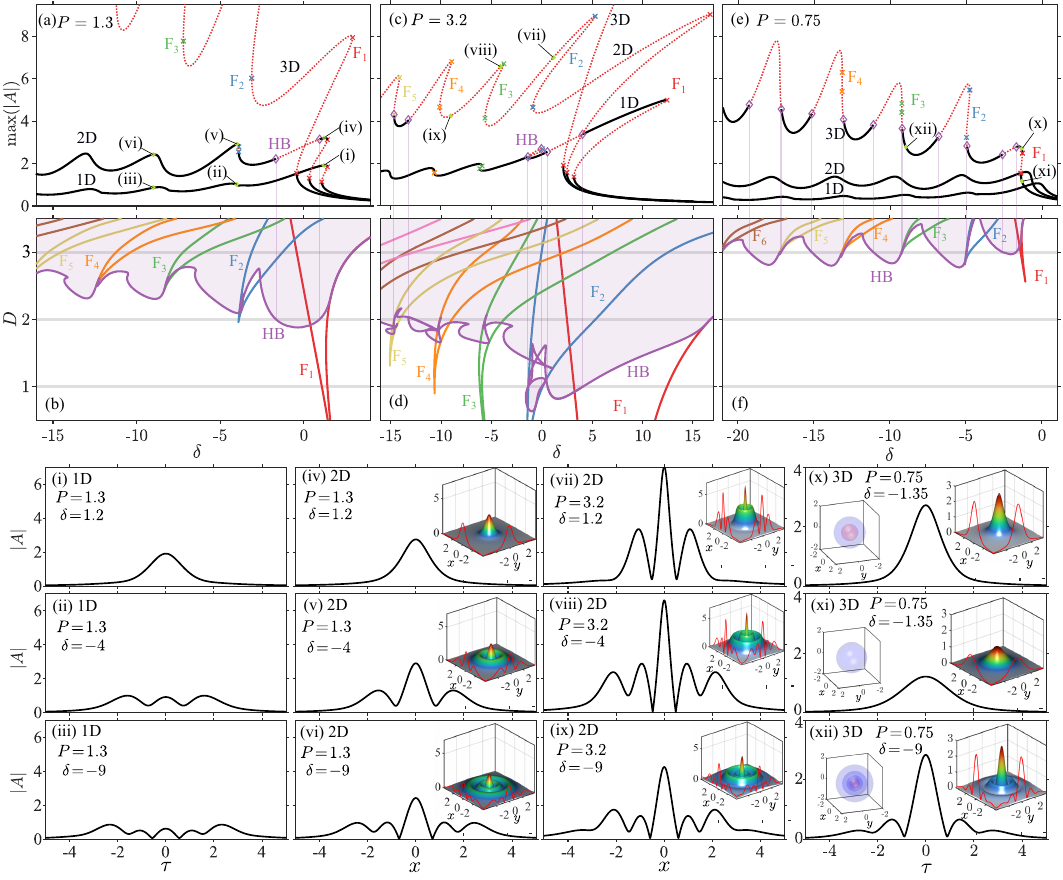}	
		\caption{
			Bifurcation diagrams for $\mathrm{max}(A)$ vs. $\delta$ and phase diagrams in the $(D, \delta)$ plane for the three parameter values: $P=1.3$ in (a,b), $P=3.2$ in (c,d), and $P=0.75$ in (e,f). In (a,c,e), stable solutions are shown as solid black curves, while unstable solutions are represented by red dashed curves. In (b,d,f), purple curves correspond to Hopf bifurcations (HB), with the remaining curves indicating fold bifurcations (FB) $F_{n}$. 
   (i-xii) Individual solutions for the field amplitude distribution $|A|$, corresponding to the solution marked solutions (i-iii) for 1D and (iv-vi) for 2D in diagram (a), (vii-ix) for 2D in diagram (c), and (x-xii) for 3D in diagram (e).
		}
		\label{fig4_pha}
	\end{figure*}
 
	Building on the previous findings, we perform a much more detailed bifurcation analysis of the LS for a continuous variation of the dimension parameter. 
    For enhancing the comparison between different dimensions, we represent the diagram of the nonlinear LS shown in Figs.~\ref{fig3_non}(a, b, c) using the maximum amplitude $\mathrm{max}(|A|)$ in the bifurcation diagram presented in Fig.\ref{fig4_pha}(a).

	The LS solutions (i-vi) in Fig.\ref{fig4_pha}(a) are illustrated in Figs.~\ref{fig4_pha}(i-vi), through the extended $r$-profiles at $\delta=1.2,\,-4,\,-9$ for 1D and 2D, respectively. 
	Subpanels in Figs.~\ref{fig4_pha}(iv-vi) depict the 2D reconstructions of such profiles.
	%{\color{blue}We have confirmed that DNS and path-continuation lead to identical resuls and stability regimes. For a more detailed discussion regarding this comparison we recommend the interested reader to consult Appendix \ref{appen:compare_DNS_conti_eigenvalue}.}
	The (i-vi) 1D and 2D states are stable, while 3D LS are mainly unstable [see Fig.\ref{fig4_pha}(a)]. Indeed, higher dimensions exhibit broader unstable and coexistence regions.
	These regions can be numerically tracked by path-continuing the fold and Hopf bifurcations in $D$ and $\delta$. As a result, we obtained the $(D,\delta)$-phase diagram of Fig.~\ref{fig4_pha}(b), which offers important insights into the dynamical behavior of the system across the different dimensions. 
	
	In Fig.\ref{fig4_pha}(b), within each FB curve $\mathrm{F}_m$, where $m=1,2,3,...$, we observe regions of coexistence of different LS, which are generated due to the tilt of the resonance peaks.
	Apparently, as the dimensionality increases, the system exhibits different numbers of pairs of FB points: one pair for $D=1$, two pairs for $D=2$, and many more pairs when $D=3$. This phenomenon can be understood from the fact that higher dimensions concentrate much stronger field intensities around $r=0$, resulting in larger intensity-dependent phase shifts. This, in turn, leads to much more pronounced tilts of the resonance peaks.
	More importantly, it is possible to map the system stability boundaries by path-continuing the HB points in Fig.~\ref{fig4_pha}(a), involving variations in both $\delta$ and $D$. 
    The resulting HB curve, depicted in purple in Fig.\ref{fig4_pha}(b), intersects the horizontal gray line at two points corresponding exactly to the HB points in Fig.~\ref{fig4_pha}(a). 
	This curve organizes the LS according to their stability: above this curve, all states in the shallow purple region are unstable. 
	
	% We conduct a similar bifurcation analysis for higher pump amplitudes. For $P=3.2$, the bifurcations diagrams associated with three dimensions are shown in Fig.~\ref{fig4_pha}(c), while its associated $(D, \delta)$-phase diagram is presented in
	% Fig.~\ref{fig4_pha}(d).
	% As can be seen, the bifurcation diagram of 1D LS exhibits a more pronounced tilt, exhibiting four pairs of FB, and two pairs of HB. For the 2D scenario, a substantial portion of the bifurcation diagram is unstable, except for a narrow region, approximately spanning the interval $-15<\delta<-13$. 
	% Unstable LS solutions located at positions (vii,viii,ix) in Fig.\ref{fig4_pha}(c) correspond to the profiles shown in 
	% Fig.\ref{fig4_pha}(vii-ix). These states are more strongly localized around the origin $r=0$ than those of  Fig.\ref{fig4_pha}(iv-vi). 
 %    The 3D diagram in Fig.~\ref{fig4_pha}(c) resembles the 2D case, but the LS has a significantly higher peak intensity.

We perform a comparable bifurcation analysis for a larger pump. Specifically, for $P=3.2$, the bifurcation diagrams corresponding to three dimensions are depicted in Fig.\ref{fig4_pha}(c), and its associated $(D, \delta)$-phase diagram is presented in Fig.\ref{fig4_pha}(d).
In contrast to the previous case illustrated in Figs.\ref{fig4_pha}(a), the bifurcation diagram in Figs.\ref{fig4_pha}(c) displays a more pronounced tilt, featuring four pairs of FB in 1D and two pairs of HB. In the 2D scenario, a considerable portion of the bifurcation diagram is unstable, except for a confined region spanning approximately the interval $-15<\delta<-13$.
The unstable 2D LS (vii, viii, ix) in Fig.\ref{fig4_pha}(c) correspond to the profiles depicted in Fig.\ref{fig4_pha}(vii-ix). These states exhibit stronger localization around the origin $r=0$ compared to those with the same detunings in Fig.\ref{fig4_pha}(iv-vi). The 3D diagram in Fig.\ref{fig4_pha}(c) mirrors the 2D case, but the LS exhibits a significantly higher peak intensity.

Importantly, the phase diagram in Figs.~\ref{fig4_pha}(d) provides a clearer view, revealing that an increase in pump amplitude results in a more complex dynamic map. The fold bifurcation not only expands its region in detuning $\delta$ but also extends further into a larger region in lower dimensions. 
The unstable region, indicated by the purple HB curves, enlarges its domain towards lower dimensions, reaching even $D=1$.

	% By looking at Fig.~\ref{fig4_pha}(d), one can see that high-order FB extend over larger $\delta$-regions with decreasing $D$, reaching even $D=1$. By path-continuing the HB on the slope of the first resonance in 1D [see Fig.~\ref{fig4_pha}(c)], the HB purple curve in ig.~\ref{fig4_pha}(d) is obtained, which permits to assess the temporal unstable regions for different dimensions at once. Although this Hopf curve cross the $D=1$ and $D=2$ lines, it does not connect with the HB emerging from the slope of the second resonance peak in the 1D LS-related diagram shown in Fig.~\ref{fig4_pha}(c), which in contrast, forms an island in the $(D, \delta)$-phase diagram.

	%By tracing this second HB point, a circular curve is obtained in Fig.~\ref{fig4_pha}(d), within which, the states are breathers. 
	%This indicating that further increasing pump would lead to more unstable regions.
	
	For lower $P$ values, the coexistence and breathing regions shrink, as shown in the bifurcation and phase diagrams of Fig.~\ref{fig4_pha}(e, f) for $P=0.75$.
	Here, the stable 3D LS corresponding to positions (x,xi,xii) in Fig.~\ref{fig4_pha}(e) are shown in Fig.\ref{fig4_pha}(x,xi,xii), along with their 2D and 3D representations. At such a low pump value, all nonlinear states in 1D and 2D are stable. In Fig.\ref{fig4_pha}(f), only the initial four FB curves intersect at $D=3$ whereas the Hopf bifurcation (HB) curves exhibit periodic intersections, creating a periodic alternation between stable and unstable intervals in Fig.~\ref{fig4_pha}(e). These results are in-line with our previous results \cite{Sun2023bullets}.

	\section{Discussions and conclusions}\label{sec:6}
	We would like to emphasize the key advantages, as well as the limitations of the proposed approach. 
	Firstly, considering radially symmetric solutions for reducing high-dimensional systems to lower ones permits to reduce complexity, which permits to gain physical insight, and to improve numerical accuracy thanks to the massively decreases computational time. 
	Secondly, treating the system dimension $D$ as a continuous and controllable parameter allows for a seamless exploration of solutions across their dimension, greatly facilitating the understanding of stability changes when transitioning from 1D to 3D structures. 
	In the context of pattern forming systems, this has been proven to be extremely useful for analyzing the modification LS and their associated homoclinic snaking toward dimensions in the Swift-Hohenberg equation \cite{lloyd_localized_2009,mccalla_snaking_2010}.
	
	Nevertheless, reducing the problem complexity also leads to some trade-offs. 
	In cases with a relatively weak pump, the radial symmetric simplification yields results which are virtually identical to those obtained in their full dimensional problems. 
	However, as the pump intensity increases, a greater variety of instabilities may emerge. The radial symmetric simplification might remove any possibility of studying less constrained scenarios, involving more morphologically complex states.  
	Consequently, in situations with a relatively high pump, the stability of the LS, derived via path-continuation of the radial profiles, may necessitate additional confirmations through DNS or linear stability analysis performed for the original problem across its full coordinates. 
 % An illustrative instance of such comparative validation in the 2D case by using DNS is given in Appendix \ref{appen:compare_DNS_conti_eigenvalue}. Even in 1D, the radial symmetry restriction may overlook other instabilities, such as those associated with asymmetric behavior \cite{Sun2023chaos}. An example of comparative validation in the 1D case, conducted through linear stability analysis in both reduced and full 1D coordinates, is presented in Appendix \ref{appen:comarisions_full}.
		
	As a matter of fact, in systems with translation invariance (i.e., in the absence of the localizing potential), states with different morphology (e.g., localized stripe, hexagons and rhomboid patches) may arise from radial LS via symmetry-breaking bifurcations \cite{lloyd_localized_2008,avitabile_snake_2010}, which are absent in our case. This is also intrinsically related to stability properties of LS. Here we only study how the system is stable against radial perturbations, thus neglecting any other degree of freedom, such as those associated with transverse or azimuthal perturbations, which are closely related to the geometric curvature of the nonlinear states \cite{gomila_stable_2001,tlidi_curvature_2002,gomila_phase-space_2007}. 

	%The effectiveness of our proposed approach relies on the inherent symmetric properties within the system's solutions. 
	%Consequently, we can reduce even a 2D or 3D problem into a one-dimensional reduced format.  
	%Secondly, our focus is specifically on evaluating the linear stability of radial field profiles. It is essential to further check that asymmetric instability may reduce the stable region. Therefore, additional investigations, such as DNS or 
	
	% To summarize, in this article we have analyzed the formation of radial dissipative LS of different dimensions in externally driven optical cavities subject to the effect of a parabolic potential. To do so, we have transformed higher-dimensional models into a single 1D model with a dimension parameter. 
 % We have performed a detailed bifurcation analysis by using the discrete dimension parameter $D$ as a continuous variable. 
 % This has permitted us to gain essential information about the connection between LS of different dimensions, and on the underlying linear eigenstates of the system. 
 % We have unveiled how the stability of radial states intrinsically depends on the system dimension. In order to illustrate that, we have drawn phase diagrams in the $(D,\delta)$-parameter space for different values of the pump strength. We have found that, as the dimension parameter $D$ increases, the system displays an expansion of the LS coexistence regions, albeit concurrently encountering a diminished stability range.

To summarize, in this article, we have analyzed the formation of radial symmetric LS of different dimensions in externally driven optical cavities subject to the effect of a parabolic potential. 
We have transformed higher-dimensional models into a single 1D model with a dimension parameter. 
This transformation provides an additional perspective: compared to the 1D system, 2D and 3D introduce an effective position-dependent field flow. 
This raised flow rate generates an effective centripetal "force" weighted by dimension parameter $D$, compelling the field to concentrate at the LS center.
Linear eigensolutions reveal heightened concentrations of eigenmodes with consistent equal spacing among eigenvalues as the dimension parameter rises. 
Furthermore, we have performed a detailed bifurcation analysis of nonlinear solutions by using the dimension parameter $D$ as a continuous variable. 
This has permitted us to gain essential information about the connection between LS of different dimensions: For the same pumping and detuning system parameters, localized states in higher-dimensional systems exhibit greater field concentration due to the introduced effective centripetal "force". 
This increased intensity results in broader bistable regions but diminished stable dynamical regions.
The results are illustrated in  $(D,\delta)$ phase diagrams for different values of the pump strength. As the dimension parameter $D$ or pump $P$ increases, the system displays an expansion of the LS coexistence regions, albeit concurrently encountering a diminished stability range. This study introduces an approach to address high-dimensional problems, illuminating the fundamental dimensional connections among radially symmetric states across varying dimensions, and supplying valuable analytical tools.

In future research, we plan to extend the present work by including a spectral stability analysis of non-radial perturbations. This will provide the detection of symmetry breaking bifurcations, and point to new ways for studying the emergence of LS with non-radially symmetric morphology.

	\begin{acknowledgements}
		This work was supported by Marie Sklodowska-Curie Actions (101064614,101023717) and Sapienza University Grants (NOSTERDIS,EFFILOCKER), and the PNRR MUR project PE0000 023-NQSTI.
	\end{acknowledgements}
	
	\bibliography{references}

\end{document}